\documentclass[pra,twocolumn,tightenlines,showpacs,amsmath,amssymb,nofootinbib]{revtex4}
\usepackage{amssymb}
\usepackage{amsmath}
\usepackage{colordvi}
\usepackage{amsthm}

\begin{document}
\newtheorem{theorem}{Theorem}
\newtheorem{lemma}{Lemma}
\newtheorem{corollary}{Corollary}

\title{Physical accessible transformations on a finite number of quantum states}
\author{Xiang-Fa Zhou}
\email{xfzhou@mail.ustc.edu.cn}
\author{Qing Lin}
\email{qlin@mail.ustc.edu.cn}
\author{Yong-Sheng Zhang}
\email{yshzhang@ustc.edu.cn}
\author{Guang-Can Guo}
\affiliation{\textit{Key Laboratory of Quantum Information,
University of Science and Technology of China, Hefei, Anhui
230026, People's Republic of China}}

\begin{abstract}
We consider to treat the usual probabilistic cloning, state
separation, unambiguous state discrimination, \emph{etc} in a
uniform framework. All these transformations can be regarded as
special examples of generalized completely positive trace
non-increasing maps on a finite number of input states. From the
system-ancilla model we construct the corresponding unitary
implementation of pure $\rightarrow$ pure, pure $\rightarrow$
mixed, mixed $\rightarrow$ pure, and mixed $\rightarrow$ mixed
states transformations in the whole system and obtain the
necessary and sufficient conditions on the existence of the
desired maps. We expect our work will be helpful to explore what
we can do on a finite set of input states.
\end{abstract}

\maketitle

\section{Introduction}
Recently, rapid progress has been made on quantum information
science. Superposition and entanglement play the central roles and
make it quite different from its classical correspondence.
Benefited from these novel properties many interesting
applications have been proposed in the past years, such as dense
coding \cite{dense}, quantum teleportation \cite{tele}, quantum
cryptography \cite{cryp}, \emph{etc}. On the other hand, quantum
superposition also puts many constraints on physical realizable
manipulation on quantum states. For example, we cannot copy (or
delete) an unknown qubit. This statement is known as the
no-cloning (no-deleting) principle and constitutes one of the most
significant differences between classical and quantum information.
Hence to get to know what we can and cannot do for a given set of
quantum states becomes an important and interesting problem.

Many efforts have been devoted into this question. One example is
the universal quantum operation (universal quantum state cloning
\cite{cloning}, universal state estimation \cite{discrimination},
\emph{etc}). In the universal quantum operation the input states
compose of the whole Hilbert space, and usually such a
transformation can be carried out determinately with the fidelity
be independent of the input states. In the most practical cases,
people often get to know some specific information about the input
states set and particularly the input usually contains only a
finite number of quantum states. Therefore the problem we focus on
in this paper is: given a finite set of input states
$\{\rho_1,\rho_2,\hdots,\rho_n\}$ (${\cal H}_1$) and the
corresponding output states
$\{\sigma_1,\sigma_2,\hdots,\sigma_n\}$ (${\cal H}_2$), we
consider whether there exists a physical accessible operation $\xi
: {\cal H}_1 \rightarrow {\cal H}_2$ to implement such kind of
transformation. Here $\rho_i$ ($\sigma_i$) denotes the input
(output) state density matrix, and it can be pure or mixed. We
also require the operation should be accurate and probabilistic.
Many interesting things can be enclosed in the framework, such as
probabilistic cloning \cite{probablistic}, state separation
\cite{separation}, unambiguous states discrimination
\cite{unambiguous}, \emph{etc}, and actually they are only some
particular cases of general completely positive (CP) maps between
two finite sets of input states.

Usually any physical evolution can be described by a CP trace
non-increasing map. There are many other equivalent ways to
illustrate such transformation. For example, it can be represented
in an operator sum form $\xi(\rho)=\sum_k M_k \rho M^\dag_k$ where
$M_k$ are the Kraus operators and satisfy $\sum_k M^\dag_k M_k \le
I$ \cite{kraus}. Also it can be implemented by employing a unitary
transformation on the extended Hilbert space. The corresponding
mathematical description can be written as
$\xi(\rho)=\mbox{Tr}_{E'}[U \rho \otimes \rho_E U^\dag I \otimes
P_{E'}]$ \cite{dueng}, where $\rho \in {\cal H}_1$, ${\cal H}_1
\otimes {\cal H}_E = {\cal H}_2 \otimes {\cal H}_{E'}$,  $\rho_E$
is the initial state  of the ancilla, $I$ denote the identity
operator in output Hilbert space ${\cal H}_2$, and $P_{E'}$ is a
projector in ${\cal H}_{E'}$. In our consideration we choose the
second kind of description since it is easy understandable and has
been widely used in many related works \cite{probablistic}.

The paper is organized as follows. In section II, under some
simple assumptions we consider the sufficient and necessary
conditions about the existence of the CP map $\xi$. We divide the
whole question into four parts, that is pure $\rightarrow$ pure,
pure $\rightarrow$ mixed, mixed $\rightarrow$ pure, and mixed
$\rightarrow$ mixed states cases. From the system-ancilla model we
obtain the sufficient and necessary conditions on the existence of
physical accessible operations. We also examine the consequences
of these conditions under some special examples, such as
probabilistic cloning of linear independent quantum states, state
separation, unambiguous state discrimination, \emph{etc.} Compared
with the method used in already existing results, we think our
results are more apparent and easier understandable. In section
III, we consider the influence of the auxiliary system to whole
problem, which has been neglected in some related works
\cite{probablistic}, and propose a generalized probabilistic
cloning machine. We also give a brief discussion of its
relationship to standard semidefinite programming. We conclude our
remarks in section VI.

\section{CP trace non-increasing map on a finite set of input states}
In this section, we try to find the sufficient and necessary
conditions on the existence of the map $\xi$. Generally this
question is quite complicated. To simplify our consideration, we
assume that the initial state of the auxiliary system is a pure
state and divide the whole problem into four parts. We think this
will be helpful to understand the main results of this paper.

\subsection{CP trace non-increasing map between two pure states sets}
\label{sec1} Let us begin with the pure-state to pure-state case,
where the input and output states are both pure and can be
represented as $\xi:\{|\varphi_1 \rangle, |\varphi_2
\rangle,\cdots, |\varphi_n \rangle \} \rightarrow \{ |\phi_1
\rangle, |\phi_2 \rangle,\cdots, |\phi_n \rangle \}$. This kind of
transformation has been
 studied in many related works, for example, probabilistic cloning
 and state separation, where the input states are required to be
 linear independent. However, for a general pure-state to
 pure-state transformation, this requirement can be loosen.
 Additionally if there exists a quantum operation $\xi$ to
 complete such kind of map, one can always construct
 the following unitary transformation
 \begin{eqnarray} \label{2.1.1}
 U|\varphi_i \rangle_1 |0\rangle_E= \sqrt{\eta_i}|\phi_i \rangle_2 |\alpha_i \rangle_a | P_0
 \rangle_p
 + | \tilde{\beta}_i \rangle_{2ap},     \{ 1 \leq i  \leq n \}, \label{eq:1}
 \end{eqnarray}
where ${\cal H}_{E'}={\cal H}_{a} \otimes {\cal H}_{p}$,
$|P_0\rangle$ is the state of the probe system satisfying $\langle
P_0|\tilde{\beta}_i\rangle=0$, and $\eta_i$ is the success
probability. Here and in the following, we use the tilde
$\tilde{\mbox{ }}$ to denote a nonnormalized state vector and omit
the subscriptions for simplicity. Since any unitary transformation
preserves inner-product, we can easily obtain $X=\sqrt{\Gamma}Y
\sqrt{\Gamma^\dag} \circ A+B$ with $X_{ij}=\langle
\varphi_i|\varphi_j \rangle$, $\sqrt{\Gamma}=diag\{\sqrt{\eta_1},
\sqrt{\eta_2}, \ldots \}$, $Y_{ij}=\langle \phi_i|\phi_j \rangle,
A_{ij}=\langle \alpha_i|\alpha_j \rangle$, and $B_{ij}=\langle
\tilde{\beta}_i|\tilde{\beta}_j \rangle$. It is not difficult to
demonstrate that the two matrices $A$ and $B$ are both positive
semidefinite. Thus we obtain
\begin{eqnarray} \label{2.1.2}
X - \sqrt{\Gamma} Y \sqrt{\Gamma^\dag} \circ A=B \geq 0,
\label{eq:2}
\end{eqnarray}
where ``$\circ$'' denotes the Hadamard (or Schur) matrix product.

On the other hand, if one can find an efficiency matrix $\Gamma$
and a positive semidefinite matrix $A$ satisfying Eq.
($\ref{eq:2}$), we can construct a unitary transformation to
realize Eq. ($\ref{eq:1}$). Since
$X-\sqrt{\Gamma}Y\sqrt{\Gamma^\dag}\circ A$ is Hermitian and
positive semidefinite, it can be diagonalized by unitary
transformation
\begin{eqnarray}
V[X-\sqrt{\Gamma}Y\sqrt{\Gamma^\dag} \circ
A]V^\dag&=&diag\{b_1,b_2,\cdots,b_n\}, \nonumber \\
&\mbox{ }&\forall m, b_m \geq 0.
\end{eqnarray}
By choosing $n$ orthogonal states $\{ |1\rangle, |2\rangle,
\cdots, |n\rangle \}$ satisfying $\langle P_0|k\rangle=0$
appropriately and setting $|\tilde{\beta}_i\rangle=\sum_{k=1}^m
V_{ki}\sqrt{b_k}|k\rangle$, we can check that $\langle
\tilde{\beta}_i|\tilde{\beta}_j \rangle=\sum_k V^\dag_{ik} b_k
V_{kj}=[X-\sqrt{\Gamma}Y\sqrt{\Gamma^\dag} \circ A]_{ij}$. Hence
with standard Gram-Schmidt process the desired unitary map can be
constructed to satisfy Eq. ($\ref{eq:1}$).

We conclude the above discussion by the following lemma.
\begin{lemma}\label{lem:1}
For any given state secretly chosen from the set
$\{|\varphi_1\rangle, \cdots, |\varphi_n\rangle \}$, it can be
transformed to the corresponding output state in $\{ |\phi_1
\rangle,\cdots, |\phi_n\rangle \}$ with the success probability
matrix $\Gamma=diag\{\eta_1,\cdots,\eta_n \}$ if and only if there
exists a positive semidefinite matrix $A$ with the diagonal
elements $A_{ii}=1$ such that $X-\sqrt{\Gamma}Y\sqrt{\Gamma^\dag}
\circ A \geq 0$. If the input states are chosen with prior
probability $\{ p_1,\cdots,p_n \}$ and $\sum_i p_i=1$, then the
whole success probability will be $P=\sum_i p_i\eta_i$.
\end{lemma}

This lemma characterizes the general property of physical
accessible maps between pure states. Usually it does not give any
constraints on the input states, that means $\{ |\varphi_1
\rangle, \cdots, |\varphi_n \rangle \}$ can be linear dependent or
linear independent. However, when the output states are linear
independent, the matrix $Y$ will be full-rank $rank(Y)=rank(Y
\circ A)=n$. In order to implement the CP map with non-zero
efficiency $\eta_k
>0$, the matrix $X$ also must be full-rank, this leads to the
linear independence of the input states. Many interesting
questions can be enclosed into this case, for example,
probabilistic copy and state discrimination of pure states. Since
the output states are linear independent, the transformation
become possible only when the input states are also linear
independent.

As a special example, in the following we concentrate on
deterministic transformation between pure states. This problem has
been considered by Chefles \cite{determinate} using different
method, while in our framework, it becomes more apparent and
easier understandable. Since $\Gamma$ is identity matrix,
according to our lemma, we obtain
\begin{eqnarray} \label{aa}
X- Y \circ A =0 \Longleftrightarrow \langle \varphi_i|\varphi_j
\rangle = \langle \phi_i|\phi_j \rangle \langle \alpha_i|\alpha_j
\rangle.
\end{eqnarray}
Eq. ($\ref{aa}$) tells us that the modules of the overlap of the
initial states should be less than their final correspondence. If
one can find a positive semidefinite matrix $A$ with $A_{ii}=1$
satisfying $X = Y \circ A$, a deterministic transformation between
the two state sets can be realized. Additionally, if $\langle
\phi_i|\phi_j \rangle \neq 0$, the matrix $A$ can be easily
constructed as $A_{ij} = \langle \varphi_i|\varphi_j
\rangle/\langle \phi_i|\phi_j \rangle$. Therefore by judging
whether $A$ is positive semidefinite or not, we can get to know
the existence of a deterministic transformation.

In the case of unambiguous discrimination between pure states, the
output states are orthogonal to each other. This leads to a
simplified version of Eq.($\ref{2.1.2}$)
\begin{eqnarray}
X - \Gamma \geq 0.
\end{eqnarray}
We then conclude that \emph{$N$ pure states can be unambiguous
discriminated if and only if there exists an efficiency matrix
$\Gamma$ such that $X-\Gamma$ is positive semidefinite}, which has
been discussed in many related works
\cite{probablistic,separation}.

\subsection{CP trace non-increasing map between pure states and mixed states}
\label{sec2} Now we continue our consideration by assuming the
output states to be mixed $\{ \sigma_1, \sigma_2, ... \sigma_n
\}$. This indicates that we should find the map
$\xi(|\varphi_i\rangle \langle \varphi_i|)=\sigma_i$ with
$\sigma_i$ be mixed state. Since the initial state of the
auxiliary system is a fixed pure state, from the system-ancilla
model we can obtain that the transformation can always be written
as follows
\begin{eqnarray}
U|\varphi_i\rangle_1 |0\rangle_E =
\sqrt{\eta_i}|\phi_i\rangle_{2a}|P_0\rangle_p +
|\tilde{\beta}_i\rangle_{2ap}.
\end{eqnarray}
The output states can be obtained by tracing out the subsystem $a$
after a measurement on the probe, \emph{i.e.}
$\sigma_i=\mbox{Tr}_a(|\phi_i\rangle \langle \phi_i|)$. Therefore
$|\phi_i\rangle_{2a}$ is nothing but a purification of the output
state $\sigma_i$. Thus by introducing the purification of the
output mixed state, the question we considered now can be reduced
to the former case where both of the input and output states are
pure. This leads to the following lemma.
\begin{lemma}\label{lem:2}
For any given state secretly chosen from the set
$\{|\varphi_1\rangle, \cdots, |\varphi_n\rangle \}$, it can be
transformed to the corresponding mixed output state in $\{
\sigma_1 ,\cdots, \sigma_n \}$ with the success probability matrix
$\Gamma=diag\{\eta_1,\cdots,\eta_n \}$ if and only if
$X-\sqrt{\Gamma}Y\sqrt{\Gamma^\dag} \geq 0$, where $Y_{ij}=\langle
\phi_i|\phi_j\rangle$ with $| \phi_i \rangle$ be the purification
of the output state $\sigma_i$. Additionally if the input states
are chosen with prior probability $\{ p_1,\cdots,p_n \}$ and
$\sum_i p_i=1$, then the whole success probability will be
$P=\sum_i p_i\eta_i$.
\end{lemma}

Since for any given mixed state, there are infinite types of
purified states. What's more, the purified states are usually
entangled states. Therefore the question we considered now becomes
quite different from the former case. For example, in the case of
deterministic transformation, one has $X=Y$, which means there
exists a set of purified states $\{ |\phi_1 \rangle, \cdots,
|\phi_n \rangle \}$ such that the corresponding
inter-inner-product equals to the case of input states. This can
be regarded as a generalized version of pure-state to pure-state
case, since $|\phi_i\rangle |\alpha_i\rangle$ is also a
purification of $|\phi_i\rangle$. However one cannot obtain a
corresponding positive semidefinite matrix $A$ easily because a
purified state of $\sigma_i$ is usually an entangled state.

The requirement of the input states to be linear independent is
also different now. Generally the linear independency of
$\sigma_i$ cannot assert that the corresponding purifications are
also linear independent. But if the support of any output state is
not contained in the combinations of the supports of the rest
mixed states, that is $supp(\sigma_i) \nsubseteq \bigoplus_{k \neq
i}^n supp(\sigma_k)$, then any purified states $\{|\phi_i \rangle
\}$ are linear independent, and $Y$ is an invertible positive
definite matrix. From the lemma, we obtain that the input states
must be linear independent.

In the case of two input states ($n=2$), the condition in lemma 2
can be written as
\begin{eqnarray}
\left(
\begin{array}{ll}
1-\eta_1 & \langle \varphi_1|\varphi_2 \rangle-\sqrt{\eta_1 \eta_2}
\langle \phi_1|\phi_2 \rangle \\
\langle \varphi_2|\varphi_1 \rangle-\sqrt{\eta_1 \eta_2} \langle
\phi_2|\phi_1 \rangle & 1- \eta_2
\end{array}
\right ) \geq 0. \nonumber
\end{eqnarray}
This implies
\begin{eqnarray}
\sqrt{(1-\eta_1)(1-\eta_2)} \geq |\langle \varphi_1|\varphi_2
\rangle| - \sqrt{\eta_1 \eta_2} |\langle \phi_1|\phi_2 \rangle|.
\end{eqnarray}
A simple algebra leads us to the following bound of the whole
success probability
\begin{eqnarray} \label{eq:3}
P = p_1 \eta_1 + p_2 \eta_2 &\leq& \frac{1-2\sqrt{p_1p_2}|\langle
\varphi_1|\varphi_2 \rangle|}{1-|\langle \phi_1|\phi_2 \rangle|}
\nonumber \\ &\leq& \frac{1-2\sqrt{p_1p_2}|\langle
\varphi_1|\varphi_2 \rangle|}{1-F(\sigma_1,\sigma_2)}.
\end{eqnarray}
Here $F(\sigma_1,\sigma_2)$ is the fidelity of the two output
state which is defined as
$F(\sigma_1,\sigma_2)=\mbox{Tr}(\sqrt{\sqrt{\sigma_1}\sigma_2
\sqrt{\sigma_1}})$. The last step of the Eq. ($\ref{eq:3}$) comes
from the well-known result that
$F(\sigma_1,\sigma_2)=\mbox{max}_{\{|\phi_1\rangle,|\phi_2\rangle\}}|\langle
\phi_1|\phi_2 \rangle|$, where the maximization runs over all
purified states \cite{nielsen,uhlmann}.

\subsection{CP trace non-increasing map between mixed states and pure states}
\label{sec3} In the above discussion, we assume that the input
states are all pure states. Here and in the following, we consider
what happens if the input set is composed of mixed states. This
problem becomes a little complicated now. And as a simple example,
we first consider the mixed-state to pure-state case $\xi: \{
\rho_1, \rho_2, \cdots, \rho_n \} \rightarrow \{ |\phi_1 \rangle,
|\phi_2 \rangle, \cdots, |\phi_n \rangle \}$.

For any mixed input state $\rho_i$, it has a spectral
decomposition
\begin{eqnarray}
\rho_i = \sum_k |\tilde{\varphi}^{(i)}_{k} \rangle \langle
\tilde{\varphi}^{(i)}_{k} |,
\end{eqnarray}
where the non-normalized state $|\tilde{\varphi}^{(i)}_{k} \rangle
= \sqrt{r^{(i)}_{k}} |\varphi^{(i)}_{k} \rangle (\parallel
|\tilde{\varphi}^{(i)}_{k} \rangle
\parallel =\sqrt{r^{(i)}_{k}} )$ are orthogonal to each other (one can also choose other kinds of decompositions
without affecting the generality of the problem we consider here).
If $supp(\rho_i) \cap supp(\rho_j) = \varnothing$ for any $i$ and
$j$, we can obtain the following equations
\begin{eqnarray}\label{mix-pure1}
U |\tilde{\varphi}^{(i)}_{k} \rangle |0\rangle
&=&\sqrt{\eta^{(i)}_{k}} |\phi_i \rangle |\alpha^{(i)}_{k} \rangle
|P_0\rangle + |\tilde{\beta}^{(i)}_{k} \rangle.
\end{eqnarray}
Here $|\alpha^{(i)}_{k} \rangle$ are arbitrary states of subsystem
$a$. If the intersection of the supports of the two density
matrices $\rho_i$ and $\rho_j$ is not empty, there exists at least
one state vector $|\psi\rangle \in supp(\rho_i) \cap
supp(\rho_j)$. From the definition of the CP map, we have
\begin{eqnarray}
U |\psi\rangle |0\rangle &=& \sqrt{\eta^{(i)}} |\phi_i\rangle
|\alpha_i\rangle|P_0\rangle + |\tilde{\beta_i}\rangle \nonumber \\
&=& \sqrt{\eta^{(j)}} |\phi_j\rangle |\alpha_j\rangle|P_0\rangle +
|\tilde{\beta_j}\rangle.
\end{eqnarray}
When the output states are \emph{different} from each other, i.e.,
$|\phi_i\rangle \neq |\phi_j \rangle$, the above equation becomes
possible only when $\eta_i=\eta_j=0$. Therefore any state with its
support contained in $supp(\rho_i) \cap supp(\rho_j)$ has no
contribution to the desired transformation. Thus in this case, it
is enough to consider $supp(\rho_i) \cap supp(\rho_j)=
\varnothing$. Otherwise, if $|\phi_i\rangle = |\phi_j \rangle$,
the components in $supp(\rho_i) \cap supp(\rho_j)$ cannot be
neglected.

Eq.($\ref{mix-pure1}$) leads to
\begin{eqnarray} \label{mix-pure}
\tilde{X}- \sqrt{\Gamma} Y \sqrt{\Gamma^\dag} \circ A \geq 0
\end{eqnarray}
with
\begin{eqnarray}
w=
\left ( \begin{array}{ccc} w_{ii} & \ldots &  w_{ij} \\
\vdots & \ddots & \vdots
\\  w_{ji} & \cdots & w_{jj}
\end{array} \right )  \,\,\,\, \{w \in (\tilde{X}, Y, A) \},
\end{eqnarray}
where $(\tilde{X}_{ij})_{kl}=\langle
\tilde{\varphi}^{(i)}_{k}|\tilde{\varphi}^{(j)}_{l} \rangle$,
$(A_{ij})_{kl}=\langle \alpha^{(i)}_{k}|\alpha^{(j)}_{l} \rangle$,
$(Y_{ij})_{kl}=\langle \phi_i|\phi_j \rangle$, and
$\Gamma=diag\{\Gamma_i, \cdots, \Gamma_j
\}=diag\{diag\{\eta^{(i)}_{1},\eta^{(i)}_{2},\cdots
\},\cdots,diag\{\eta^{(j)}_{1},\eta^{(j)}_{2},\cdots \} \}$.
$\tilde{X}_{ij}$ arises from the decomposition of input mixed
states and $A_{ij}$ is determined by the auxiliary subsystem $a$.
Accordingly we arrive at the following lemma
\begin{lemma}\label{lem:3}
For any given mixed state chosen from the set $\{\sigma_i, \cdots,
\sigma_n \}$, it can be transformed respectively to the
corresponding pure output state $\{ |\phi_1\rangle, \cdots,
|\phi_n \rangle \}$ with the success probability matrix $\Gamma$
if and only if Eq. ($\ref{mix-pure}$) is satisfied. Moreover, if
the input states are chosen with prior probabilities $\{
p_1,\cdots,p_n \}$ with $\sum_i p_i=1$, then the whole success
probability will be $P=\sum_i p_i \mbox{Tr}(\Gamma_i)$.
\end{lemma}

For any mixed state, it can be regarded as an ensemble of pure
states. Eq. ($\ref{mix-pure}$) means that the transformation from
mixed input states to pure states is equivalent to transformation
from pure states sets to pure states, i.e. $\xi:
\{\{|\tilde{\varphi}^{(i)}_{1} \rangle, |\tilde{\varphi}^{(i)}_{2}
\rangle, \cdots \},\cdots, \{|\tilde{\varphi}^{(j)}_{1} \rangle,
|\tilde{\varphi}^{(j)}_{2} \rangle, \cdots \} \} \rightarrow \{
|\phi_i\rangle, \cdots, |\phi_j\rangle \}$, which makes the
question a little different from the former cases. For example, we
consider the deterministic transformation from mixed-state to
pure-state. This indicates the efficiency matrix $\Gamma_i=diag\{
\sqrt{r^{(i)}_{1}}, \sqrt{r^{(i)}_{2}}, \cdots\}$. Now Eq.
($\ref{mix-pure}$) can be written as
\begin{eqnarray} \nonumber
\left ( \begin{array}{ccc} X_{ii} & \cdots &  X_{ij} \\
\vdots & \ddots & \vdots
\\  X_{ji} & \cdots & X_{jj}
\end{array} \right )
= \left ( \begin{array}{ccc} A_{ii} & \cdots & \langle
\phi_i|\phi_j \rangle A_{ij} \\ \vdots & \ddots & \vdots
\\ \langle \phi_j|\phi_i \rangle A_{ji} & \cdots & A_{jj}
\end{array} \right )
\end{eqnarray}
with $(X_{ij})_{kl}=\langle \varphi^{(i)}_{k}|\varphi^{(j)}_{l}
\rangle$. Hence \emph{if one can find such a positive semidefinite
matrix $A$ ( with its diagonal elements $(A_{ii})_{kk}=1$) to
ensure that the above equation is satisfied, we can implement the
desired transformation determinately.} If $\langle \phi_i|\phi_j
\rangle \neq 0$, one can construct the matrix $A$ very easily
$A_{ij}= X_{ij}/\langle \phi_1|\phi_2 \rangle$. By judging the
positive definiteness of $A$, we can get to know the existence of
a deterministic transformation. As a simple case, we assume there
are only two input states ($\rho_1$ and $\rho_2$) considered here.
In the case of pure input states, deterministic transformation
exists only when $\langle \varphi_1|\varphi_2 \rangle \leq
F(\sigma_1, \sigma_2)$. While in the mixed input states case, the
above equation can be simplified into
\begin{eqnarray}
\left ( \begin{array}{ll} I_{11} &  X_{12}/\langle \phi_1|\phi_2 \rangle  \\
 X_{21}/\langle \phi_2|\phi_1 \rangle  & I_{22}
\end{array} \right )
= \left ( \begin{array}{ll} A_{11} &  A_{12} \\  A_{21} & A_{22}
\end{array} \right ) .
\end{eqnarray}
Here $I_{11}$, $I_{22}$ are both identity matrices. Thus $A$ is
positive semidefinite if and only if $X_{12}X_{21} \leq |\langle
\phi_1|\phi_2 \rangle|^2 I_{11}$. This means the maximal singular
value of $X_{12}$ which arises from the normalized eigenvectors of
the two mixed input states must not be larger than the fidelity of
the two output states $|\langle \phi_1|\phi_2 \rangle|$.

\subsection{CP trace non-increasing map between two mixed states sets}
\label{sec4} In the general case, the input and output states are
both mixed. Since for a mixed state $\rho_i=\sum_k
|\tilde{\varphi}^{(i)}_{k} \rangle \langle
\tilde{\varphi}^{(i)}_{k}| =\sum_m |\tilde{\psi}^{(i)}_{m} \rangle
\langle \tilde{\psi}^{(i)}_{m}|$, it can be generated from very
many different kinds of ensembles. This will make the question
more complicated. Following the same routine as before, we can
write down the unitary implementation of the CP map
\begin{eqnarray}
U |\tilde{\varphi}^{(i)}_{k} \rangle |0\rangle
&=&|\tilde{\phi}^{(i)}_{k} \rangle |P_0\rangle +
|\tilde{\beta}^{(i)}_{k} \rangle,
\end{eqnarray}
and the output states $|\tilde{\phi}^{(i)}_{k} \rangle$ should
satisfy $\sum_k \mbox{Tr}_a(|\tilde{\phi}^{(i)}_{k} \rangle
\langle \tilde{\phi}^{(i)}_{k} |) = \eta_i \sigma_i$, where
$\eta_i$ describes the success probability of the CP map on
$\rho_i$. When the output or the input states are pure, one can
check that $|\tilde{\phi}^{(i)}_{k} \rangle$ are proportional to
the purifications of the output states. Also if the output states
are different from each other and their supports have no common
nonzero vectors, we can obtain the states lying in $supp(\rho_i)
\cap supp(\rho_j)$ have no contribution the transformation.

Similarly, the inner-product preservation of unitary
transformation lead us to
\begin{eqnarray} \label{mix-mix}
\left ( \begin{array}{ccc} \tilde{X}_{ii} & \cdots &  \tilde{X}_{ij} \\
\vdots & \ddots & \vdots
\\  \tilde{X}_{ji} & \cdots & \tilde{X}_{jj}
\end{array} \right )
-  \left ( \begin{array}{ccc} \tilde{Y}_{ii} & \cdots &
\tilde{Y}_{ij}
\\ \vdots & \ddots & \vdots
\\ \tilde{Y}_{ji} & \cdots & \tilde{Y}_{jj}
\end{array} \right )  \geq 0,
\end{eqnarray}
where $(\tilde{Y}_{ij})_{kl}=\langle
\tilde{\phi}^{(i)}_{k}|\tilde{\phi}^{(j)}_{l} \rangle$. Therefore
we have
\begin{lemma}\label{lem:4}
The transformation $\xi$ from mixed states to mixed state $\xi:
\{\rho_i, \cdots, \rho_j, \cdots \} \rightarrow \{ \sigma_i,
\cdots, \sigma_j, \cdots \}$ can be implemented with the success
probabilities $\{\eta_i, \cdots, \eta_j, \cdots \}$ if and only if
there exists sets of composite states
$\theta_i=|\tilde{\phi}^{(i)}_{k} \rangle \langle
\tilde{\phi}^{(i)}_{k} |$ with $\sum_k \mbox{Tr}_a(\theta_i) =
\eta_i \sigma_i$ such that Eq. ($\ref{mix-mix}$) is satisfied. If
the input states are chosen with prior probability $\{ p_i,
\cdots, p_j, \cdots \}$ and $\sum_i p_i=1$, then the whole success
probability will be $P=\sum_i p_i \eta_i=\sum_i p_i
\mbox{Tr}(\tilde{Y}_{ii})$.
\end{lemma}

Lemma $\ref{lem:4}$ characterizes the most general properties of
CP maps between mixed states. By assuming the output or the input
states are pure, we can immediately obtain the corresponding
results in section ($\ref{sec1}-\ref{sec3}$). To make our result
more specific, let us focus on an interesting case where the
output states $\{ \sigma_i \}$ are all orthogonal to each other.
This question is a very special case of transformation between
mixed states and has drawn much attention recently (unambiguous
discrimination of mixed states) \cite{two-mixed}. Since $\sigma_i
\perp \sigma_j$, we have $\langle \tilde{\phi}^{(i)}_{k}
|\tilde{\phi}^{(j)}_{l} \rangle=0$ $(i\neq j)$ for any $k$ and
$l$. This implies that the matrix $\tilde{Y}$ is quasi-diagonal
and can be expressed as $\tilde{Y}=diag\{\tilde{Y}_{ii}, \cdots,
\tilde{Y}_{jj}, \cdots \}$. Hence we obtain that \emph{$N$ mixed
states  $\{\rho_1, \rho_2, \cdots \}$ can be unambiguously
discriminated if and only if there exists a positive semidefinite
quasi-diagonal matrix $\tilde{Y}=diag\{\tilde{Y}_{11},
\tilde{Y}_{22}, \cdots \}$ such that $\tilde{X}- \tilde{Y} \geq
0$}.

As another interesting example, in the following we assume there
are only two input states contained here. In this case, Eq.
($\ref{mix-mix}$) can be rewritten as
\begin{eqnarray}
\left ( \begin{array}{ccc} \tilde{X}_{11} & \tilde{X}_{12} \\
 \tilde{X}_{21} & \tilde{X}_{22}
\end{array} \right )
-  \left ( \begin{array}{ccc} \tilde{Y}_{11} & \tilde{Y}_{12} \\
\tilde{Y}_{21} & \tilde{Y}_{22}
\end{array} \right )  \geq 0.
\end{eqnarray}
Without loss of generality, we also suppose both $\tilde{X}_{11}$
and $\tilde{X}_{22}$ are $t \times t$ matrices. From the standard
linear algebra theory, we obtain the following inequalities
\begin{eqnarray}
\sqrt{(r^{(1)}_k-\eta^{(1)}_k)(r^{(2)}_k-\eta^{(2)}_k)} \geq
|\langle \tilde{\varphi}^{(1)}_k|\tilde{\varphi}^{(2)}_k \rangle -
\langle \tilde{\phi}^{(1)}_k|\tilde{\phi}^{(2)}_k \rangle |.
\end{eqnarray}
It's not difficult to find that
\begin{eqnarray}
p_1 \eta_1 + p_2 \eta_2  \leq \frac{1-2\sqrt{p_1p_2}\sum_k
|\langle \tilde{\varphi}^{(1)}_k|\tilde{\varphi}^{(2)}_k \rangle
|}{1- \sum_k |\langle \tilde{\phi}^{(1)}_k|\tilde{\phi}^{(2)}_k
\rangle|/\sqrt{\eta_1 \eta_2}}.
\end{eqnarray}

Since different ensembles can give rise to the same mixed states,
this is known as the unitary freedom for density matrices, we have
\begin{eqnarray} \label{P}
P \leq \frac{ 1- 2 \sqrt{p_1 p_2} \,
\mbox{max}_{\{|\tilde{\varphi}^{(1)}_k \rangle,
|\tilde{\varphi}^{(2)}_k \rangle \} } \sum_k |\langle
\tilde{\varphi}^{(1)}_k|\tilde{\varphi}^{(2)}_k \rangle |} {1-
\mbox{max}_{\{|\tilde{\phi}^{(1)}_k \rangle, |\tilde{\phi}^{(2)}_k
\rangle \} } \sum_k |\langle
\tilde{\phi}^{(1)}_k|\tilde{\phi}^{(2)}_k \rangle|/\sqrt{\eta_1
\eta_2}},
\end{eqnarray}
where $P=p_1 \eta_1 + p_2 \eta_2$, and the maximization is over
all decompositions of the input mixed states satisfying
$\rho_i=\sum_k | \tilde{\varphi}^{(i)}_k \rangle \langle
\tilde{\varphi}^{(i)}_k|$. Interestingly, the right hand side of
Eq. ($\ref{P}$) is directly related to the fidelity of the two
input mixed states $F(\rho_1,
\rho_2)=\mbox{Tr}\sqrt{\sqrt{\rho_1}\rho_2\sqrt{\rho_1}}$. To show
this, we introduce the purification of $\rho_1$ and $\rho_2$
\begin{eqnarray}
|\varphi_1 \rangle = \sqrt{\rho_1} U^{(1)} \otimes U^{(1)}_Q
|m\rangle,
\nonumber \\
|\varphi_2 \rangle = \sqrt{\rho_2} U^{(2)} \otimes U^{(2)}_Q
|m\rangle, \nonumber
\end{eqnarray}
where $|m\rangle=\sum_i |i\rangle |i_Q\rangle$ is, up to a
normalization factor, a maximally entangled state of the input
system and the extended system Q (we assume that the two systems
have the same rank) \cite{nielsen}. Choosing
$U^{(1)}_Q=U^{(2)}_Q$, one can find suitable unitary matrices
$U^{(1)}$ and $U^{(2)}$ such that
\begin{eqnarray}
\langle \varphi_1|\varphi_2 \rangle = \sum_k |\langle
\tilde{\varphi}^{(1)}_k|\tilde{\varphi}^{(2)}_k \rangle |.
\end{eqnarray}
From the Uhlmann's theorem \cite{nielsen,uhlmann} we have
\begin{eqnarray}
 F(\rho_1, \rho_2) &=& \mbox{max}_{\{|\varphi_1 \rangle, |\varphi_2 \rangle \}}\langle
\varphi_1|\varphi_2 \rangle \nonumber \\  &=&
\mbox{max}_{\{|\tilde{\varphi}^{(1)}_k \rangle,
|\tilde{\varphi}^{(2)}_k \rangle \} } \sum_k |\langle
\tilde{\varphi}^{(1)}_k|\tilde{\varphi}^{(2)}_k \rangle |.
\nonumber
\end{eqnarray}
Thus Eq. ($\ref{P}$) now can be simplified as
\begin{eqnarray}
P &\leq& \frac{ 1- 2 \sqrt{p_1 p_2} F(\rho_1, \rho_2)} {1-
\mbox{max}_{\{|\tilde{\phi}^{(1)}_k \rangle, |\tilde{\phi}^{(2)}_k
\rangle \} } \sum_k |\langle
\tilde{\phi}^{(1)}_k|\tilde{\phi}^{(2)}_k \rangle|/\sqrt{\eta_1
\eta_2}} \nonumber \\ &\leq& \frac{1-2 \sqrt{p_1 p_2} F(\rho_1,
\rho_2)}{1-F(\sigma_1, \sigma_2)}.
\end{eqnarray}
This result is actually the generalize version of Eq.
($\ref{eq:3}$). One can check that when $\sigma_1 \perp \sigma_2$,
 $P \leq 1 - 2 \sqrt{p_1p_2} F(\rho_1, \rho_2) $, which is also consistent
 with already known results on unambiguous discrimination between two mixed states \cite{two-mixed}.

\section{The role of auxiliary system and semidefinite programming}
In the above sections, we have considered the condition under
which a physical accessible transformation exists on a finite
number of input states. By constructing the unitary realization of
the CP maps, we have presented the sufficient and necessary
conditions on the existence of the desired transformation. The
auxiliary system $a$ plays a very important role in our
consideration. Usually in order to implement a CP transformation
it is necessary to introduce such a subsystem. For instance, in
the deterministic transformation between pure states, if $|\langle
\varphi_1|\varphi_2 \rangle| < |\langle \phi_1|\phi_2 \rangle|$,
it is impossible to complete the transformation $\{|\varphi_1
\rangle, |\varphi_2 \rangle \} \rightarrow \{|\phi_1 \rangle,
|\phi_2 \rangle \}$ without the ancilla system $a$.

As we have mentioned before, probabilistic cloning can be regarded
as a special transformation on a finite number of linear
independent states. However, in the original work
\cite{probablistic}, the subsystem $a$ has been neglected, and the
corresponding unitary map is defined as
\begin{eqnarray}
U|\varphi_i \rangle |0\rangle = \sqrt{\eta_i} |\varphi_i \rangle
^{\otimes N} |P_0\rangle + |\tilde{\beta}_i\rangle.
\end{eqnarray}
This is equivalent to saying that probabilistic cloning can be
implemented by setting all $|\alpha_i \rangle$ in Eq.
($\ref{2.1.1}$) to be equal, or equivalently $A$ is a rank-1
operator. Generally it is difficult to say this since
$X-\sqrt{\Gamma} Y \sqrt{\Gamma^\dag} \circ A \geq 0$ can not
ensure $X-\sqrt{\Gamma} Y \sqrt{\Gamma^\dag} \geq 0$. This also
indicates that in some cases we can find a probability matrix
$\Gamma$ which satisfies $X-\sqrt{\Gamma} Y \sqrt{\Gamma^\dag}
\ngeq 0$, i.e., it cannot be implemented by the probabilistic copy
machine in \cite{probablistic}, but can be realized in our
framework if one can find a suitable matrix $A$. In this sense, we
define the generalized probabilistic cloning machine as
\begin{eqnarray}
U|\varphi_i \rangle |0\rangle = \sqrt{\eta_i} |\varphi_i \rangle
^{\otimes N} |\alpha_i \rangle |P_0\rangle +
|\tilde{\beta}_i\rangle.
\end{eqnarray}
However, if the input set contains only two states, the optimal
probabilistic copy machine can be implemented without the ancilla
system $a$ \cite{probablistic}.

In a more realistic situation, people often concentrate on the
whole success probability of such physical transformation. This
indicates that we should make the probability $P$ as high as
possible. Interestingly, if the subsystem $a$ is contained and if
we know exactly the two positive semidefinite matrices $X$ and $Y$
(this happens when the output states are pure or orthogonal to
each other), we can reduce this to a standard semidefinite program
(SDP) problem \cite{sdp}.

Usually a standard SDP problem is to maximize
\begin{eqnarray}
\sum_m b_m y_m
\end{eqnarray}
subject to
\begin{eqnarray}
C-\sum_{i=1}^n y_m S_m \geq 0,
\end{eqnarray}
where $b\in R^{n}$, $y \in R^{n}$ with $b$ be a given vector,
$b_m$, $y_m$ are the $m$th components of vector $b$ and $y$
respectively, and $C$ and $S_m$ are given Hermitian matrices.
While in our framework we should maximize (for example, see Lemma
1)
\begin{eqnarray} \label{sdp0}
P=\sum_k p_k  \tilde{A}_{kk}
\end{eqnarray}
under the constraints
\begin{eqnarray} \label{sdp2}
X-Y \circ \tilde{A} \geq 0, \, and \, \tilde{A} \geq 0.
\end{eqnarray}
Here we have assumed
$\tilde{A}=\sqrt{\Gamma}A\sqrt{\Gamma^{\dag}}$ for simplicity. Now
we define
\begin{eqnarray}
F_0 &=& \left(
\begin{array}{cc}
X & 0 \\
0 & 0
\end{array}
\right ), \nonumber \\
F_{kl} &=& \left(
\begin{array}{ll}
Y_{kl} E_{kl} + Y_{lk} E_{lk} & 0 \\
0 & -E_{kl} -E_{lk}
\end{array}
\right ), \nonumber \\
G_{kl} &=& i \left(
\begin{array}{ll}
Y_{kl} E_{kl} - Y_{lk} E_{lk} & 0 \\
0 & - E_{kl} + E_{lk}
\end{array}
\right ), \nonumber
\end{eqnarray}
where $i$ is the basic imaginary unit with $i^2=-1$, $Y_{kl}$
denotes the ($k$, $l$)th entry of the matrix $Y$, and $E_{kl}$ are
matrices with all entries be zero except the ($k$, $l$)th elements
$(E_{kl})_{mn}=\delta_{km} \delta_{ln}$. Eq. ($\ref{sdp0}$) can be
reformulated as
\begin{eqnarray}
{\hbox{\space \raise-2mm\hbox{$\textstyle max \atop \scriptstyle
\tilde{A}$} \space}} \sum_k p_k \tilde{A}_{kk}
\end{eqnarray}
subject to
\begin{eqnarray} \label{sdp}
F_0 &-& \sum_{k,l > k} \left [ Re(\tilde{A}_{kl}) F_{kl} +
Im(\tilde{A}_{kl}) G_{kl} \right ] \nonumber \\ &-&
\frac{1}{2}\sum_k \tilde{A}_{kk} F_{kk} \geq 0,
\end{eqnarray}
where $Re(\tilde{A}_{kl})$ and $Im(\tilde{A}_{kl})$ represent the
real and imaginary part of $\tilde{A}_{kl}$ respectively. Thus,
this problem has been reduced to a standard SDP problem, which can
be solved numerically to $\epsilon$-optimization in polynomial
time. It can be found that the subsystem a plays a very important
role in the reducing step, which ensures the elements
$\tilde{A}_{kl}$ to be independent of each other. Otherwise, we
cannot reduce the whole problem to SDP easily. Moreover, to judge
the existence of the matrix $A$ now is equivalent to determining
where there exist $\tilde{A}_{kl}$ to make Eq. ($\ref{sdp}$)
satisfied. This is often called semidefinite feasibility problem
(SDFP). Unfortunately, this problem has not be totally solved yet
\cite{sdp}.

As a specific example, consider the case where the input contains
three states with equal prior probability, i.e., $| \varphi_1
\rangle = \left( 2|0\rangle + |1\rangle + |2\rangle \right
)/\sqrt{6}$, $|\varphi_2\rangle = \left( |0\rangle + 3|1\rangle +
|2\rangle \right )/\sqrt{11}$, and $|\varphi_3\rangle = \left(
|0\rangle + |1\rangle + 4|2\rangle \right )/(3\sqrt{2})$. We
concentrate on a general state separation problem (generalized
version of probabilistic cloning), that is to transform these
input states into the following output states $| \phi_1 \rangle
=(10|0\rangle + |1\rangle + |2\rangle)/\sqrt{101}$, $| \phi_2
\rangle =(|0\rangle + 10 |1\rangle + |2\rangle)/\sqrt{101}$, and
$| \phi_3 \rangle =(|0\rangle + |1\rangle + 10
|2\rangle)/\sqrt{101}$ separately. The optimal solution can be
found when
\begin{eqnarray}
A= \left (
\begin{array}{ccc}
0.1570 & 0.2329 & 0.2643 \\
0.2329 & 0.4342 & 0.2977 \\
0.2643 & 0.2977 & 0.5452
\end{array}
\right ).
\end{eqnarray}
This matrix has two non-zero eigenvalues with $\lambda_1=0.0$,
$\lambda_2=0.1874$, and $\lambda_3=0.9491$. The corresponding
success probability matrix is $\Gamma=diag\{ 0.1570, 0.4342,
0.5453 \}$. We can also check that $X-\sqrt{\Gamma}Y\sqrt{\Gamma}$
is not a positive semidefinite matrix, hence the corresponding
transformation cannot be implemented without the subsystem $a$,
which agrees with the above presentation.

\section{conclusion}
In summary, we have considered physical accessible transformations
on a finite set of input states. From the system-ancilla model, by
constructing the appropriate unitary map, we obtain the sufficient
and necessary conditions on the existence of the desired
transformation. Many interesting questions can be enclosed in this
framework, such as deterministic transformation between quantum
states, probabilistic cloning of linear independent states, state
separation, unambiguous state discrimination, \emph{etc}. Our
discussion reveals that all these problems can be treated
uniformly, and actually they are only special cases of general CP
maps on a finite input states set. In the practical viewpoint,
usually people have only a finite number of different states in
hand, therefore to explore what kinds of operations we can do and
how to do on these quantum resources becomes a very important
problem. We expect our results will be useful in judging the
existence of the CP maps and also helpful in constructing these
transformations.

We thank Y.-J. Han and Y.-C. Wu for helpful discussions. This work
was funded by the National Fundamental Research Program , the
National Natural Science Foundation of China (Grants No. 10304017
and No. 60121503), the Innovation Funds from the Chinese Academy
of Sciences, and Program for New Century Excellent Talents in
University.
\end{document}